\documentclass[a4paper]{jpconf}

\usepackage[utf8]{inputenc}
\usepackage[T1]{fontenc}
\usepackage{graphicx}
\usepackage{hyperref}
\usepackage{url}
\usepackage{amsmath,amssymb,amsthm}
\usepackage{color}
\usepackage{enumitem}
\usepackage{tikz}
\usetikzlibrary{arrows}
\usetikzlibrary{shapes}
\usetikzlibrary{calc}
\usetikzlibrary{decorations.pathreplacing}

\newtheorem{theorem}{Theorem}%[section]

\newtheorem{proposition}{Proposition}

\newtheorem{definition}{Definition}
\newtheorem{remark}{Remark}

\newcommand{\paren}[1]{\left(#1\right)}
\newcommand{\brac}[1]{\left[#1\right]}

\newcommand{\norm}[1]{\left\|#1\right\|}
\newcommand{\set}[1]{\left\{#1\right\}}
\newcommand{\abs}[1]{\left\lvert #1 \right\rvert}

\newcommand{\wt}[1]{\widetilde{#1}}

\newcommand{\order}[1]{\mathcal{O}\paren{#1}}

\def \Exp {\mathbb{E}}
\def\PP{\mathbb{P}}

\def \imaginaryi {\textrm{i}}

\DeclareMathOperator*{\argmin}{arg\,min}

\def \RR {\mathbb{R}}
\def \CC {\mathbb{C}}
\def \NN {\mathbb{N}}

\def \Model {\mathfrak{S}}
\def\secant{\mathcal{S}}
\def \freq {{\boldsymbol \omega}}

\def \pFail {\rho}
\def \Decoder {\Delta}

\newcommand{\covnum}[3]{\mathcal{N}\left(#1,#2,#3\right)}

\def \coveps {\delta}

\def \ve {\mathbf{e}}
\def \vx {\mathbf{x}}

\def \mI {\mathbf{I}}

\def \AnyOpNA{\Psi}
\newcommand{\AnyOp}[1]{\Psi\paren{#1}}
\def \ProjNA{\textup{P}_\Model}
\newcommand{\Proj}[1]{\textup{P}_\Model\paren{#1}}
\def \AnySig {\vx}
\def \AnyMeas {\mathbf{y}}
\def \AnySigSet {E}
\def \AnyMeasSpace {F}

\newcommand{\normAnyMeas}[1]{\norm{#1}_\AnyMeasSpace}
\newcommand{\normAnySig}[1]{\norm{#1}_\AnySigSet}
\def \AnyNoise {\ve}
\def \IOPCstA {A}
\def \IOPCstB {B}
\def \IOPMetricA {d_\AnySigSet}
\def \IOPMetricB {d_\AnySigSet'}
\def \LRIPCst {\alpha}
\def \LRIPMetric {d_\AnySigSet}
\def \BPMetric {d_{G}}
\def \BPCst {\beta}
\def \LRIPError {\eta}

\def \DecodError {\lambda}
\newcommand{\tsig}[1]{{#1^\star}}

\newcommand{\recov}[1]{\wt{#1}}

\definecolor{darkpurple}{rgb}{0.3,0,0.3}

\usepackage{amsthm}

\begin{document}
\title{Instance Optimal Decoding and the Restricted Isometry Property}
\author{Nicolas Keriven}
\address{\'Ecole Normale Supérieure. 45 rue d'Ulm, 75005 Paris, France.}
\ead{nicolas.keriven@ens.fr}
\author{R\'emi Gribonval}
\address{Universit\'e Rennes 1, Inria, CNRS, IRISA. F-35000 Rennes, France.}
\ead{remi.gribonval@inria.fr}

\begin{abstract}
%This short article summarizes some results that previously appeared as a chapter in \cite{thesis} but were not published independently.
In this paper, we address the question of \emph{information preservation} in ill-posed, non-linear inverse problems, assuming that the measured data is close to a low-dimensional model set. We provide necessary and sufficient conditions for the existence of a so-called \emph{instance optimal} decoder, $i.e.$, that is robust to noise and modelling error. Inspired by existing results in compressive sensing, our analysis is based on a (Lower) \emph{Restricted Isometry Property} (LRIP), formulated in a non-linear fashion. We also provide sufficient conditions for non-uniform recovery with random measurement operators, with a new formulation of the LRIP. We finish by describing typical strategies to prove the LRIP in both linear and non-linear cases, and illustrate our results by studying the invertibility of a one-layer neural net with random weights.
\end{abstract}

\section{Introduction}

%%% general intro
Inverse problems are ubiquitous in all areas of data science. While \emph{linear} inverse problems have been arguably far more studied in the literature, some frameworks are intrinsically non-linear \cite{Engl2005}. In this paper, we aim at giving a characterization of the \emph{preservation of information} in ill-posed inverse problems, regularized by the introduction of a ``low-dimensional'' \emph{model set} close to which the data of interest is assumed to live. We consider a very general context that includes, in particular, measurement operators that are possibly non-linear, and an ambient space that can be any pseudometric set. Our main results show that the existence of a decoder that is robust to noise and modelling error is \emph{equivalent} to a modified \emph{Restricted Isometry Property} (RIP), which is a classical property in compressive sensing \cite{Candes2008}. We thus outline the fundamental nature of the RIP in settings that are more general than previously studied.

%%% notation, problem
The problem is formulated as follows. Let $(\AnySigSet,\IOPMetricA)$ be a set equipped with a pseudometric\footnote{A pseudometric $d$ satisfies all the requirements of a metric except $d(x,y) = 0 \Rightarrow x=y$.} $\IOPMetricA$, the set of data, and $(\AnyMeasSpace,\normAnyMeas{\cdot})$ a seminormed\footnote{Similarly, a seminorm satisfy the requirements of a norm except that $\norm{x}=0$ does not imply $x=0$.} vector space, the space of measurements. Consider a (possibly non-linear) measurement map $\AnyOpNA:\AnySigSet \to\AnyMeasSpace$. The measured vector is:
\begin{equation}
\label{eq:sig_meas}
\AnyMeas = \AnyOp{\tsig{\AnySig}} + \AnyNoise
\end{equation}
where $\AnyNoise \in \AnyMeasSpace$ is measurement noise and $\tsig{\AnySig}$ is the true signal. Our goal is to characterize the existence of \emph{any} procedure that would allow us to approximately recover the data $\tsig{\AnySig}$ from $\AnyMeas$.

%%% strategy, goal
\paragraph{Regularization.} In most interesting problems, the ``dimension'' of the space $\AnyMeasSpace$ is far lower than that of the set $\AnySigSet$ (in a loose sense: we recall that here none is required to be finite-dimensional, and $\AnySigSet$ is not necessarily a vector space), which makes the problem \emph{ill-posed}, meaning that there are information-theoretic limits that prevent us from recovering the underlying signal from the measurements, even in the noiseless case. A classical regularization technique is to introduce \emph{prior knowledge} about the true signal $\tsig{\AnySig}$, here we consider a \emph{model set} $\Model \subset \AnySigSet$ of ``simple'' signals, such that $\tsig{\AnySig}$ is likely to be close to $\Model$. For instance, sparsity, $i.e.$ the assumption that the true signal is a linear combination of a few elements in a well-chosen dictionary, is a hugely-studied prior in modern signal processing, in particular in compressive sensing \cite{Foucart2013}.

\paragraph{Instance Optimal Decoding.} Ideally, a \emph{decoder} $\Decoder:\AnyMeasSpace \to \AnySigSet$ must be able to exactly retrieve $\tsig{\AnySig}$ from $\AnyMeas$ when the modelling is exact ($i.e.$ $\tsig{\AnySig}\in\Model$) and the noise is zero ($\AnyNoise = 0$). However, as these conditions are highly unrealistic in practice, it is desirable for this decoding process to be both robust to noise and stable to \emph{modelling error}. In the literature, such a decoder is said to be \emph{instance optimal} \cite{Cohen2009} (see Def. \ref{def:IOP}). In this paper, our goal is to characterize necessary and sufficient conditions for the existence of an instance optimal decoder for the problem \eqref{eq:sig_meas}. Note that we will not study the existence of efficient algorithms to solve \eqref{eq:sig_meas}, which is another significant achievement of compressive sensing \cite{Foucart2013}, but only the \emph{preservation of information} of the encoding process.

In \cite{Cohen2009,Bourrier2014a}, the authors outlined the crucial role played by the \emph{Restricted Isometry Property} (RIP), and more precisely by the \emph{Lower}-RIP (LRIP), for the existence of instance optimal decoders in the linear case. %, which roughly states that the measurement operator $\AnyOpNA$ approximately preserves distances between elements in the model.
In this paper, we extend these results to the non-linear case and to non-uniform probabilistic recovery. %, when the operator $\AnyOpNA$ is drawn at random.

%As we will see, the existence of such decoder is equivalent to the so-called \emph{Lower Restricted Isometry Property}, simply written in the non-linear and generic metrics case. Furthermore, when the operator $\AnyOpNA$ is drawn at random, one can make an additional distinction between \emph{uniform} and \emph{non-uniform} recovery, and still derivate IOP and LRIP results in this case.

%%% outline
\paragraph{Outline of the paper.} The structure of the paper is as follows. In Section \ref{sec:related} we briefly outline some relevant references, keeping in mind that the field is large and we do not pretend to be exhaustive in this short paper. In Section \ref{sec:main} we state our main results relating instance optimal decoders (Def. \ref{def:IOP}) and the LRIP (Def. \ref{def:LRIP}). In Section \ref{sec:proof_LRIP} we outline how one might typically prove the LRIP by extending a classical proof \cite{Baraniuk2008}, and illustrate it on a simple example. % that involves random features for kernel approximation.

\section{Related Work}\label{sec:related}

%Both the RIP and instance optimal decoding are now well-established notions, particularly in the field of Compressive Sensing. 

%In this section we briefly give some relevant references, 

\paragraph{Classical Compressive Sensing: the linear case.} Instance optimal decoding and the RIP are well-known notions in compressive sensing \cite{Candes2004,Candes2006b,Donoho2006}. We refer to the book of Foucart and Rauhut \cite{Foucart2013} for a review of the field, in particular to Chapters 6 and 11 for the topics of interest here. The interplay between the two notions was in particular studied in \cite{Cohen2009} in the finite-dimensional case. These results were later extended to more general models in \cite{Peleg2013}, and to any linear measurement operators in \cite{Bourrier2014a}, which is the main inspiration behind the present work.

\paragraph{Non-uniform decoding.} In compressed sensing the measurement operator $\AnyOpNA$ is often designed at random. Typical recovery results are therefore given \emph{with high probability}. One can then distinguished between \emph{uniform} guarantees, meaning that with high probability on the draw of $\AnyOpNA$ all signals $\tsig{\AnySig}$ close to $\Model$ can be stably recovered, and non-uniform guarantees, $i.e.$ for one \emph{fixed} signal $\tsig{\AnySig}$ close to $\Model$, with high probability on $\AnyOpNA$ the decoding is successful. In \cite{Cohen2009} the authors study non-uniform instance optimality, but only under the light of the classical uniform RIP. In this paper we introduce a non-uniform version of the LRIP and prove that it is sufficient for non-uniform instance optimality.

\paragraph{Non-linear inverse problems.} Non-linear inverse problems can be found in many areas of signal processing, see e.g. \cite{Engl2005} for a review of some applications. They have also been considered by the compressive sensing community, often when \emph{quantization} occurs \cite{Jacques2015,Boufounos2015a}, in the so-called ``1-bit'' compressed sensing line of work \cite{Boufounos2008}.
Another focus is the development of efficient algorithms inspired by the linear case \cite{Beck2012,Blumensath2013}. In \cite{Blumensath2013}, the author assume that a locally linearized version of $\AnyOpNA$ satisfy the classical RIP. In this paper we consider a different, ``fully'' non-linear RIP. We note that one notion does not imply the other.

%\paragraph{Contribution.} Inspired by the results in \cite{Bourrier2014a}, we prove that the existence of an instance optimal decoder is equivalent to the LRIP in our general settings.

\section{Equivalence between IOP and LRIP}\label{sec:main}

In this section we state our main results on instance optimal decoders and the LRIP. We distinguish the case where the operator $\AnyOpNA$ is deterministic, or, equivalently, when it is random but one seeks so-called \emph{uniform} recovery guarantees, and the case of non-uniform recovery.

\subsection{Deterministic operator}

%%% recall problem
Recall that we consider a pseudometric set $(\AnySigSet,\IOPMetricA)$, a seminormed vector space $(\AnyMeasSpace,\normAnyMeas{\cdot})$, and measurements of the form $\AnyMeas = \AnyOp{\tsig{\AnySig}} + \AnyNoise$ where $\AnyOpNA:\AnySigSet \to\AnyMeasSpace$. We consider a model set $\Model \subset \AnySigSet$, and are interested in characterizing the existence of a good \emph{decoder} $\Decoder(\AnyOpNA,\AnyMeas)$ that takes $\AnyOpNA$ and $\AnyMeas$ as inputs and return a signal $\recov{\AnySig} \in \AnySigSet$ that is close to $\tsig{\AnySig}$.  We want this decoder to be stable to modelling error and robust to noise, which is characterized by the notion of instance optimality.

\begin{definition}[Instance Optimality Property (IOP)]\label{def:IOP}
A decoder $\Decoder$ satisfies the Instance Optimality Property for the operator $\AnyOpNA$ and model $\Model$ with constants $\IOPCstA,\IOPCstB >0$, pseudometrics $\IOPMetricA, \IOPMetricB$ on $\AnySigSet$ and error $\DecodError\geq 0$ if: for all signals $\tsig{\AnySig}\in\AnySigSet$ and noise $\AnyNoise \in \AnyMeasSpace$, denoting $\recov{\AnySig}=\Decoder(\AnyOpNA, \AnyOp{\tsig{\AnySig}} + \AnyNoise)$ the recovered signal, it holds that:
\begin{equation}
\label{eq:IOP}
\IOPMetricA(\tsig{\AnySig}, \recov{\AnySig}) \leq \IOPCstA \IOPMetricB(\tsig{\AnySig},\Model) + \IOPCstB \normAnyMeas{\AnyNoise} + \DecodError
\end{equation}
where $\IOPMetricB(\AnySig,\Model) = \inf_{\AnySig_\Model \in \Model}\IOPMetricB(\AnySig,\AnySig_\Model)$.
\end{definition}

%We want the returned signal to be close to the \emph{best approximation} of $\tsig{\AnySig}$ by an element in $\Model$.
As indicated by the r.h.s. of \eqref{eq:IOP}, the decoding error between the true signal and the recovered one is bounded by the amplitude of the noise and the distance from $\tsig{\AnySig}$ to the model set, which indicates how well $\tsig{\AnySig}$ is modelled by $\Model$. An instance optimal decoder is therefore robust to noise and stable even if $\tsig{\AnySig}$ is not exactly in the model set. % that the returned signal is close to the \emph{best approximation} of $\tsig{\AnySig}$ by an element in $\Model$ with respect to some metric $\IOPMetricB$. 
We also include a possible fixed additional error $\DecodError\geq 0$, which may be unavoidable in some cases (due to algorithmic precision for instance). Ideally, one has $\DecodError=0$.

%%% LRIP
%\paragraph{The Lower Restricted Isometry Property.}
Let us now turn to the proposed non-linear version of the LRIP. As described in \cite{Bourrier2014a}, the LRIP is just one side of the classical RIP, which states that the measurement operator $\AnyOpNA$ approximately preserves distances between elements of the model $\Model$.

%The Restricted Isometry Property (RIP) is a classical property in compressive sensing sparse recovery. It states that, on the model set, the (traditionally linear) measurement operator behaves almost like an isometry, in the sense that it approximately preserves the distance between two elements in the model set. We simply reformulate this notion here in the non-linear case, and only on one side of the equation (hence ``Lower'' RIP, or LRIP). The RIP has been a useful notion in proving many important results on the success of convex relaxation of sparse regularizer. Here we examine its links with the existence of \emph{any} instance optimal decoder (even potentially intractable).

%The LRIP is formulated as follows.

\begin{definition}[Lower Restricted Isometry Property (LRIP)]\label{def:LRIP}
The operator $\AnyOpNA$ satisfies the Lower Restricted Isometry Property for the model $\Model$ with constant $\LRIPCst >0$, pseudometric $\LRIPMetric$ and error $\LRIPError\geq 0$ if: for all $\AnySig,\AnySig' \in \Model$ it holds that
\begin{equation}\label{eq:LRIP}
\LRIPMetric(\AnySig,\AnySig') \leq \LRIPCst \normAnyMeas{\AnyOp{\AnySig} - \AnyOp{\AnySig'}} + \LRIPError.
\end{equation}
\end{definition}
The LRIP expresses the fact that $\AnyOpNA$ must not collapse two elements of the model together. Like the IOP, we allow for a possible additional fixed error $\LRIPError\geq 0$ in the LRIP. Note that this type of error is often considered when introducing quantization \cite{Boufounos2015a,Jacques2015}. Ideally, one has $\LRIPError=0$, however in some cases it can be considerably simpler to prove that the LRIP holds with a non-zero $\LRIPError$ \cite{Keriven2017a}. The reader would note that the classical RIP is often expressed with a constant $\alpha = (1-t)^{-1}$ where $t<1$ is a small as possible.

%. As we will illusrate later, the LRIP may sometimes be considerably simpler to prove in the case where this additional error is non-zero. In the next theorem, we see that it simply carries in the instance optimality error. 

%%% main thm
We now state our main result. The proof, rather direct, can be found in \ref{app:LRIPIOP}.

\begin{theorem}[Equivalence between IOP and LRIP.]\label{thm:LRIPIOP}
Consider an operator $\AnyOpNA$ and a model $\Model$.
\begin{enumerate}
\item If there exists a decoder $\Decoder$ which satisfies the Instance Optimality Property for $\AnyOpNA$ and $\Model$ with constants $\IOPCstA,\IOPCstB >0$, pseudometrics $\IOPMetricA, \IOPMetricB$ and error $\DecodError\geq 0$, then the operator $\AnyOpNA$ satisfies the LRIP for $\Model$ with constant $\LRIPCst = \IOPCstB$, pseudometric $\IOPMetricA$ and error $\LRIPError = 2\DecodError$.
\item If the operator $\AnyOpNA$ satisfies the LRIP for the model $\Model$ with constant $\LRIPCst$, pseudometric $\LRIPMetric$ and error $\LRIPError\geq 0$, then the decoder $\Decoder$ defined as\footnote{In this paper we assume that the minimization problem $\argmin_{\AnySig \in \Model} \normAnyMeas{\AnyOp{\AnySig} - \AnyMeas}$ has at least one solution, for simplicity (ties can be broken arbitrarily). When this is not the case, it is possible to consider a decoder that returns any element that approaches the infimum with a fixed precision, at the expense of having this precision in the decoding error $\DecodError$, as in \cite{Bourrier2014a}.}
\begin{equation}\label{eq:decoder}
\Decoder(\AnyOpNA,\AnyMeas) = \argmin_{\AnySig \in \Model} \normAnyMeas{\AnyOp{\AnySig} - \AnyMeas}
\end{equation}
satisfies the Instance Optimality Property for the operator $\AnyOpNA$ and model $\Model$ with constants $\IOPCstA=1$ and $\IOPCstB=2\LRIPCst$, pseudometrics $\LRIPMetric$ and $\IOPMetricB$ where $\IOPMetricB$ is defined by $\IOPMetricB(\AnySig,\AnySig'):=\LRIPMetric(\AnySig,\AnySig') + 2\LRIPCst\normAnyMeas{\AnyOp{\AnySig} - \AnyOp{\AnySig'}}$, and error $\DecodError = \LRIPError$.
\end{enumerate}
\end{theorem}

Theorem \ref{thm:LRIPIOP} states that if the LRIP is satisfied, then the decoder that returns the element in the model that best matches the measurement is instance optimal, with a special metric $\IOPMetricB$. On the other hand, if \emph{some} instance optimal decoder exists, then the LRIP must be satisfied.
In other words, when the LRIP is satisfied, then we know that a negligible amount of information is lost when encoding a signal well-modeled by $\Model$. Conversely, if the LRIP is not satisfied, one has no hope of deriving an instance optimal decoder. %, and therefore checking for the LRIP should be a first step.

%Furthermore, in both implications, the LRIP or IOP are robust to additional errors, due to limited algorithmic precision for instance. % does not degrade too much 

% the ideal decoder might rarely be implementable in practice, it has a theoretical interest: if one hopes to derive any instance optimal decoder then the LRIP must be satisfied by the encoding process.

\subsection{Random operator, from uniform recovery to non-uniform recovery}

%%% Problem, goal
In the vast majority of the compressive sensing literature, the measurement process is drawn at random: for instance, in the finite dimensional case, it is an open problem to find deterministic matrices that satisfies the RIP with an optimal number of measurements (\cite{Foucart2013}, pp. 27), while on the contrary many classes of random matrices satisfy the RIP with high probability \cite{Baraniuk2008}.

A well-studied concept is that of uniform recovery guarantees, where one shows that, with high probability on the draw of $\AnyOpNA$, the LRIP holds. It follows by Theorem \ref{thm:LRIPIOP} that there is a decoder such that, with high probability on the draw of $\AnyOpNA$, all signals from $\Model$ can be stably recovered. There is also a notion of \emph{non-uniform} recovery, where one considers a decoder $\Decoder$ and wonders if, given an arbitrary signal close to $\Model$, this signal is stably recovered (with high probability on the draw of $\AnyOpNA$) from $\AnyOp{\tsig{\AnySig}} + \AnyNoise$.
%This leads to the notion of \emph{non-uniform} recovery, in which a single signal $\tsig{\AnySig}$ is recovered with high probability, as opposed to recovering all signals with high probability (so-called uniform guarantees). The RIP has mainly been studied in the uniform case, since as soon as $\AnyOpNA$ satisfies the LRIP one directly obtains uniform guarantees for all signals by Theorem \ref{thm:LRIPIOP}. 
In this section we introduce a non-uniform version of the LRIP, and show that it is a sufficient condition for the existence of a non-uniform instance optimal decoder. %We start with the notion of non-uniform instance optimality.
We start by discussing a notion of projection on the model.

\begin{remark}[Approximate projection.] As we will see, in non-uniform recovery the distance from $\tsig{\AnySig}$ to $\Model$ is replaced by the distance from $\tsig{\AnySig}$ to a particular element $\AnySig_\Model = \Proj{\tsig{\AnySig}}$, where $\ProjNA:\AnySigSet \to \Model$ is a ``projection'' function with respect to some metric $\IOPMetricB$. In full generality, it is not guaranteed that there exists $\ProjNA$ such that $\IOPMetricB(\tsig{\AnySig},\Proj{\tsig{\AnySig}}) = \IOPMetricB(\tsig{\AnySig},\Model)$, but one can always define it such that for all $\tsig{\AnySig}$, $\IOPMetricB(\tsig{\AnySig},\Proj{\tsig{\AnySig}}) \leq \IOPMetricB(\tsig{\AnySig},\Model)+\varepsilon$ for an arbitrary small $\varepsilon>0$.
\end{remark}

Let us now introduce the proposed non-uniform IOP and LRIP.

\begin{definition}[Non-uniform IOP]\label{def:NU-IOP}
A decoder $\Decoder$ satisfies the non-uniform Instance Optimality Property for the (random) mapping $\AnyOpNA$, model $\Model$ and projection function $\ProjNA$, %, signal $\tsig{\AnySig} \in \AnySigSet$ and element of the model $\AnySig_\Model \in \Model$
with constants $\IOPCstA,\IOPCstB >0$, pseudometrics $\IOPMetricA, \IOPMetricB$, probability $1-\pFail$ and error $\DecodError\geq 0$ if: %for all $\tsig{\AnySig}\in\AnySigSet$, with probability at least $1-\pFail$ on the draw of $\AnyOpNA$, for all noise $\AnyNoise \in \AnyMeasSpace$, denoting $\recov{\AnySig}=\Decoder(\AnyOpNA, \AnyOp{\tsig{\AnySig}} + \AnyNoise)$,% it holds that:
\begin{equation}
\label{eq:NU-IOP}
\forall \tsig{\AnySig}\in\AnySigSet,\quad \PP_\AnyOpNA\Big[\forall \AnyNoise \in \AnyMeasSpace,~\IOPMetricA(\tsig{\AnySig}, \recov{\AnySig}) \leq \IOPCstA \IOPMetricB(\tsig{\AnySig},\Proj{\tsig{\AnySig}}) + \IOPCstB \normAnyMeas{\AnyNoise} + \DecodError\Big]\geq 1-\rho
\end{equation}
where $\Decoder(\AnyOpNA, \AnyOp{\tsig{\AnySig}} + \AnyNoise)$ is denoted by $\recov{\AnySig}$.
\end{definition}

%In the above definition, the distance from the signal $\tsig{\AnySig}$ to the model is also ``non-uniform'': it is measured as the distance to a particular $\AnySig_\Model\in\Model$ that is also fixed in advance. If it exists, one can choose $\AnySig_\Model$ such that $\IOPMetricB(\tsig{\AnySig},\AnySig_\Model) = \min_{\AnySig\in\Model}\IOPMetricB(\tsig{\AnySig},\AnySig) = \IOPMetricB(\tsig{\AnySig},\Model)$, however in full generality one can only choose $\AnySig_\Model$ such that $\IOPMetricB(\tsig{\AnySig},\AnySig_\Model) = \IOPMetricB(\tsig{\AnySig},\Model) + \varepsilon$ for an arbitrary small $\varepsilon$.

Note that in this definition the IOP is non-uniform with respect to the data $\tsig{\AnySig}$ but uniform with respect to the noise $\AnyNoise$, meaning that with high probability on the draw of $\AnyOpNA$ the (fixed) data can be stably recovered from a measurement vector with any additive noise.

%%% NU LRIP, need for BP

\begin{definition}[Non-uniform LRIP]\label{def:NU-LRIP}
The operator $\AnyOpNA$ satisfies the non-uniform LRIP for the model $\Model$ %and a fixed element of the model $\AnySig \in \Model$, 
with constant $\LRIPCst >0$, pseudometric $\LRIPMetric$, probability $1-\pFail$ and error $\LRIPError\geq 0$ if: %for all $\AnySig\in\Model$, with probability at least $1-\pFail$ on the draw of $\AnyOpNA$, for all $\AnySig'\in\Model$ we have
\begin{equation}\label{eq:NU-LRIP}
\forall \AnySig\in\Model,\quad \PP_\AnyOpNA\Big[\forall\AnySig'\in\Model,~\LRIPMetric(\AnySig,\AnySig') \leq \LRIPCst \normAnyMeas{\AnyOp{\AnySig} - \AnyOp{\AnySig'}} + \LRIPError\Big]\geq 1-\rho.
\end{equation}
\end{definition}

This LRIP is in fact ``semi''-uniform: it is non-uniform with respect to one element $\AnySig$ but uniform with respect to $\AnySig'$. A ``fully'' non-uniform LRIP would, in fact, be almost always valid for many operators (see Section \ref{sec:proof_LRIP}), and thus probably too weak to yield recovery guarantees.

Before stating our result, let us remark that the definition of the metric $\IOPMetricB$ in Theorem \ref{thm:LRIPIOP} $(ii)$ involves the operator $\AnyOpNA$, which is potentially problematic when it is random. To solve this, \cite{Cohen2009} introduces a so-called \emph{Boundedness Property} (BP) in the classical sparse setting in finite dimension. We extend this notion in the considered context here. %which defines a non-uniform upper bound on $\normAnyMeas{\AnyOp{\AnySig}-\AnyOp{\AnySig_\Model}}$, for a fixed signal $\AnySig\in\AnySigSet$ and a fixed $\AnySig_\Model\in\Model$.

\begin{definition}[Boundedness property (BP)]\label{def:BP} The operator $\AnyOpNA$ satisfies the Boundedness Property with constant $\BPCst$, pseudometric $\BPMetric$ and probability $1-\pFail$ if: %for all $\AnySig \in \AnySigSet$ and $\AnySig_\Model\in\Model$, with probability at least $1-\pFail$ on $\AnyOpNA$, we have
\begin{equation}
\label{eq:BP}
\forall \AnySig \in \AnySigSet,~\forall\AnySig_\Model\in\Model,\quad\PP_\AnyOpNA\Big[\normAnyMeas{\AnyOp{\AnySig}-\AnyOp{\AnySig_\Model}}\leq \BPCst \BPMetric\paren{\AnySig,\AnySig_\Model}\Big]\geq 1-\rho.
\end{equation}
\end{definition}

%%% result, only one side of implication

We then have the following result, proved in \ref{app:NU-IOPLRIP}.

\begin{theorem}[The non-uniform LRIP and BP implies the non-uniform IOP]\label{thm:NU-IOPLRIP}
Consider a random operator $\AnyOpNA$. %, a model $\Model$, a signal $\tsig{\AnySig}\in\AnySigSet$ and a fixed element of the model $\AnySig_\Model \in \Model$.
Assume that:
\begin{enumerate}%[label=\roman*)]
\item the operator $\AnyOpNA$ satisfies the non-uniform LRIP for the model $\Model$ with constant $\LRIPCst >0$, pseudometric $\LRIPMetric$, probability $1-\pFail_1$ and error $\LRIPError\geq 0$~;
\item the operator $\AnyOpNA$ satisfies the non-uniform Boundedness Property with constant $\BPCst$, pseudometric $\BPMetric$ and probability $1-\pFail_2$~;
\end{enumerate}
Then, the decoder $\Decoder$ defined by \eqref{eq:decoder} satisfies the non-uniform Instance Optimality Property for the operator $\AnyOpNA$, model $\Model$ and any projection function $\ProjNA$ with constants $\IOPCstA=1$, $\IOPCstB=2\LRIPCst$, pseudometrics $\LRIPMetric$ and  $\IOPMetricB := \LRIPMetric + 2\LRIPCst\BPCst\BPMetric$, probability $1-\pFail_1-\pFail_2$ and error $\DecodError = \LRIPError$.
\end{theorem}

Compared with the result in \cite{Cohen2009}, which proves non-uniform recovery under a \emph{uniform} LRIP and the BP in the finite-dimensional case, our result holds under weaker hypotheses.

For the converse implication of Theorem \ref{thm:NU-IOPLRIP}, unlike Theorem \ref{thm:LRIPIOP}, the non-uniform IOP does not seem to directly imply the non-uniform LRIP. %, and require additional hypotheses. We leave it aside for brevity.

\section{A typical proof of the LRIP}\label{sec:proof_LRIP}
%%% intro
In this section, we outline a possible strategy to prove the LRIP, inspired by the proof for random matrices in \cite{Baraniuk2008}. %While the decoder \eqref{eq:decoder} may not be computationally implementable, proving the LRIP gives \emph{preservation of information} results.
This relatively simple proof has two steps: first, a pointwise concentration result, and second, an extension by covering arguments. For a set $S$ and a metric $d$, we denote by $\covnum{S}{d}{\coveps} \in \NN \cup \set{+\infty}$ the minimum number of balls of radius $\coveps$, with centers that belong to $S$, required to cover $S$.

\subsection{Linear case}

%\todo{example of proof just by concentration function (ref ledoux, CSL), plus normalized secant set (ref CSL, Puy, Baraniuk simple RIP)}

We start with the linear case, which follows closely the proof in \cite{Baraniuk2008}. We treat the uniform case (Def. \ref{def:LRIP}), with no error ($\LRIPError = 0$).
%
%In this section, we outline a simple proof of the LRIP in the linear, uniform case, with no quantization error (\ie $\LRIPError = 0$).
Assume $(\AnySigSet,\IOPMetricA)$ and $(\AnyMeasSpace,\normAnyMeas{\cdot})$ are both vector spaces, and that we have a random \emph{linear} operator $\AnyOpNA:\AnySigSet\to\AnyMeasSpace$ such that %Our goal is to prove the uniform LRIP, \ie that there exists a constant $\LRIPCst$ such that, with high probability on the drawing of $\AnyOpNA$, for all $\AnySig,\AnySig'\in\Model$, equation \eqref{eq:LRIP} is satisfied (here with $\LRIPError = 0$).
%Assume that $\AnyOpNA$ is constructed such that 
the following \emph{concentration} result holds:
\begin{equation}\label{eq:concentration_linear}
\forall \AnySig, \AnySig' \in \Model \text{ s.t. } \IOPMetricA(\AnySig,\AnySig')>0,\quad \PP_\AnyOpNA\paren{\tfrac{\normAnyMeas{\AnyOp{\AnySig} - \AnyOp{\AnySig'}}}{\IOPMetricA(\AnySig,\AnySig')} - 1 \leq -t } \leq e^{-c(t)}
\end{equation}
for an increasing \emph{concentration function} $c(t)>0$. Typically, the ``bigger'' the space $\AnyMeasSpace$ is ($i.e.$ the more measurements we collect), the higher the concentration function is: often, for $m$ measurements ($\AnyMeasSpace = \RR^m$ or $\CC^m$), classical concentration inequalities yield $c(t) \propto mt^2$.

This property %property is often encountered in the literature. It 
proves a ``pointwise'' (or ``fully'' non-uniform) LRIP: for two given $\AnySig,\AnySig'$, the quantity $\normAnyMeas{\AnyOp{\AnySig - \AnySig'}}$ is a good approximation of $\IOPMetricA(\AnySig,\AnySig)$ with high probability. We now invert the quantifiers by covering arguments. %: we want that with high probability, the LRIP is satisfied for \emph{all} pairs of elements of the model.
%
%Such extension from a pointwise to a uniform result is classically done by resorting to $\varepsilon$-coverings: one first covers the set of interest by a finite number of balls of controlled radius, use the pointwise result at the center of each ball along with a union bound, then extend the result to the whole set by proving that inside each small ball one does not deviate too much from the center.
%
From the formulation of the concentration \eqref{eq:concentration_linear} we see that a particular set of interest is the so-called \emph{normalized secant set} \cite{Puy2015}:
\begin{equation}\label{eq:secantlinear}
\secant = \set{\tfrac{\AnySig - \AnySig'}{\IOPMetricA(\AnySig,\AnySig')} ~|~\AnySig,\AnySig'\in\Model,~\IOPMetricA(\AnySig,\AnySig')>0} \subset \AnySigSet
\end{equation}
%It is the set of differences of elements in the model, projected onto the unit ball of $\normAnySig{\cdot}$. When the concentration property holds and the normalized secant set has finite covering numbers, one can prove the LRIP.
%Using covering arguments, one can prove the LRIP. 
The proof of the following result is in \ref{app:LRIPlinear}.
\begin{proposition}\label{prop:LRIPlinear}
Consider $0<t<1$. Assume that the concentration property \eqref{eq:concentration_linear} holds, that $\secant$ has finite covering numbers, and that for any draw of $\AnyOpNA$ and any $\mathbf{s},\mathbf{s}'\in\secant$ we have $\normAnyMeas{\AnyOp{\mathbf{s}-\mathbf{s}'}}\leq C\IOPMetricA(\mathbf{s}, \mathbf{s}')$. Set $\coveps = \frac{t}{2C}$. Define the probability of failure
\[
\rho = \covnum{\secant}{\IOPMetricA}{\coveps} \cdot e^{-c(t/2)}
\]
Then the operator $\AnyOpNA$ satisfies the uniform LRIP for the model $\Model$ with constant $\LRIPCst = (1-t)^{-1}$, metric $\LRIPMetric$, probability $1-\rho$ and error $\LRIPError = 0$.
\end{proposition}

This proof of the RIP has been used for instance in classical compressive sensing \cite{Baraniuk2008} or for random linear embeddings of Radon measures \cite{Gribonval2017}. It is also used in a constructive manner to build appropriate operators $\AnyOpNA$ in \cite{Puy2015}.

\subsection{Non-linear case}
It is possible to adapt the previous proof to non-linear operators, by distinguishing the case where $\AnySig$ and $\AnySig'$ are ``close'', for which we resort to a linearization of $\AnyOpNA$ and properties of the normalized secant set, and the case where $\AnySig$ and $\AnySig'$ are distant from each other, for which we use directly the covering numbers of the model. We treat here the non-uniform case (Def. \ref{def:NU-LRIP}). %, again with no error.
%\todo{just adaptation of previous. main interest: weird metrics. Ex: random features and kernel metrics (expect limited number of random features to be sufficient for kernel method, but future work. ref bach random features). Ref non-linear recovery (Blumensath, davies, eldar...)}

Assume again that $(\AnySigSet, \LRIPMetric)$ and $(\AnyMeasSpace, \normAnyMeas{\cdot})$ are vector spaces. %Fix an element of the model $\AnySig_\Model \in \Model$, with respect to which we want to prove the non-uniform LRIP. 
Assume that we have a random map $\AnyOpNA:\AnySigSet \to \AnyMeasSpace$ such that the concentration property \eqref{eq:concentration_linear} holds.
%\begin{equation}
%\forall \AnySig, \AnySig' \in \Model,~\mathbb{P}\paren{\frac{\normAnyMeas{\AnyOp{\AnySig} - \AnyOp{\AnySig'}}}{\LRIPMetric(\AnySig,\AnySig')} - 1 \leq -t } \leq e^{-c(t)}
%\end{equation}
%for an increasing concentration function $c(t)>0$. %Typically, the ``bigger'' the space $\AnyMeasSpace$ is (i.e. the more measurements we collect), the higher the concentration function is.
Next, suppose that there exists $\varepsilon_0>0$ such that for any fixed $\AnySig\in\Model$:
\begin{enumerate}
\item for all $\AnySig'\in\Model$ and any draw of $\AnyOpNA$, $\normAnyMeas{\AnyOp{\AnySig} - \AnyOp{\AnySig'}} \leq C_1 \LRIPMetric(\AnySig, \AnySig')$
%\item for all $\AnySig,\AnySig'\in\Model$ such that $\normAnySig{\AnySig-\AnySig'}\leq \delta_0$, we have $\LRIPMetric(\AnySig,\AnySig') \leq C_2 \normAnySig{\AnySig-\AnySig'}$
\item the model $\Model$ has finite covering numbers with respect to $\LRIPMetric$, %for all $0< \varepsilon\leq \varepsilon_0$, the ``extruded model'' $\Model_\varepsilon = \Model \backslash \set{\AnySig~|~\LRIPMetric(\AnySig_\Model,\AnySig)\leq \varepsilon}$ has finite covering numbers for the norm $\normAnySig{\cdot}$,
 and in particular it also has finite diameter $M_\Model = \sup_{\AnySig,\AnySig'\in\Model} \LRIPMetric(\AnySig,\AnySig')$. 
\item  for all $0<\varepsilon\leq \varepsilon_0$, the following version of the normalized secant set $\secant_\varepsilon := \set{\frac{\AnySig - \AnySig'}{\LRIPMetric(\AnySig,\AnySig')} ~|~\AnySig' \in \Model,~0<\LRIPMetric(\AnySig,\AnySig')\leq \varepsilon}$ has finite covering numbers. % wrt $\LRIPMetric$. %Note that, unlike the previous linear case, this "secant" set is no longer the subset of a unit ball, however it is further restricted since here we consider only the case $\LRIPMetric(\AnySig_\Model,\AnySig)\leq \varepsilon$;
\item for all $\AnySig' \in \Model$ such that $\LRIPMetric(\AnySig,\AnySig')\leq \varepsilon_0$, and any draw of $\AnyOpNA$, we have
$
\abs{\AnyOp{\AnySig} - \AnyOp{\AnySig'} - D_\AnyOpNA(\AnySig - \AnySig')}\leq C_2 \LRIPMetric(\AnySig,\AnySig')^2,
$
where $D_\AnyOpNA:\AnySigSet\to \AnyMeasSpace$ is a linear map such that for all $\mathbf{s},\mathbf{s}' \in\secant_{\varepsilon_0}$, $\normAnyMeas{D_\AnyOpNA(\mathbf{s}-\mathbf{s}')}\leq C_3 \LRIPMetric(\mathbf{s},\mathbf{s}')$.
\end{enumerate}
The following result is proved in \ref{app:nonlinLRIP}.
\begin{proposition}\label{prop:nonlinLRIP}
Assume that the properties $(i)-(iv)$ above are satisfied. Set $\varepsilon \leq \min\paren{\varepsilon_0,\frac{t}{8C_2}}$, $\coveps' \leq \frac{t}{4C_3}$ and $\coveps \leq \frac{t\varepsilon^2/(4C_1)}{\varepsilon + M_\Model}$.
Define the probability of failure
\[
\rho := \brac{\covnum{\Model}{\LRIPMetric}{\coveps} + \covnum{\secant_\varepsilon}{\LRIPMetric}{\coveps'}} \cdot e^{-c(t/2)}
\]
Then the operator $\AnyOpNA$ satisfies the non-uniform LRIP for the model $\Model$ %and the element $\AnySig_\Model$, 
with constant $\LRIPCst = (1-t)^{-1}$, metric $\LRIPMetric$, probability $1-\rho$ and error $\LRIPError = 0$.
\end{proposition}

\subsection{Illustration}

In this section we illustrate the non-linear LRIP on a simple example; that of recovering a vector from a \emph{random features embedding}, which is a random map initially designed for kernel approximation, see \cite{Rahimi2007,Rahimi2009}. Such a random embedding can be seen as a one-layer neural network with random weights, for which invertibility and preservation of information have recently been topics of interest \cite{Gilbert2017a,Giryes2015}.

Consider $\AnySigSet = \RR^d$ and define $\Model$ to be a \emph{Union of Subspaces}, which is a popular model in compressed sensing \cite{Blumensath2009a}, with controlled norm: $\Model = \bigcup_{i=1}^N (S_i\cap B_M)$ where $B_M = \set{\AnySig,~\norm{\AnySig}_2\leq M}$ 
and each $S_i$ is an $s$-dimensional subspace of $\RR^d$. As in \cite{Gribonval2017}, we choose a sampling that is a reweighted version of the original Fourier sampling for kernel approximation \cite{Rahimi2007}, for the Gaussian kernel with bandwidth $\sigma>0$. It is defined as follows: for a number of measurements $m$, the measurements space is $\AnyMeasSpace = \CC^m$, the random map is defined as $\AnyOp{\AnySig} = \frac{1}{\sqrt{m}}\brac{\varphi_{\freq_j}(\AnySig)}_{j=1}^m$ where $\varphi_\freq(\AnySig) = \frac{e^{\imaginaryi\freq_j^\top \AnySig}}{f(\freq)}$ and $\freq_j\in\RR^d$ are drawn $i.i.d.$ from $\Lambda = f(\freq)^2 \mathcal{N}(0,\sigma^{-2}\mI)$ (where $\mathcal{N}$ is a Gaussian), with $f(\freq)^2 = \frac{1}{3}\sum_{\ell=0}^2\frac{\norm{\freq}_2^{2\ell}}{\gamma_{2\ell}}$ and $\gamma_\ell = \Exp_{\freq \sim \mathcal{N}(0,\sigma^{-2}\mI)} \norm{\freq}_2^\ell = \order{\frac{d^{\ell/2}}{\sigma^\ell}}$. One can verify that $\int d\Lambda(\freq) = 1$, $i.e.$ $\Lambda$ is a valid probability distribution. The metric $\LRIPMetric$ is here the \emph{kernel metric} associated to the Gaussian kernel with bandwidth $\sigma$:
\[
\LRIPMetric(\AnySig,\AnySig') := 2\paren{1 - \exp\paren{-\tfrac{\norm{\AnySig - \AnySig'}_2^2}{2\sigma^2}}}
\]
We have the following result, which proof is in \ref{app:rf}.
\begin{proposition}\label{prop:rf}
If the number of measurements is such that
\[
m\gtrsim t^{-2} \cdot\paren{ s\cdot \log\paren{\tfrac{Md}{\sigma t}} + \log(N) + \log\paren{\tfrac{1}{\rho}}}
\]
Then the operator $\AnyOpNA$ satisfies the non-uniform LRIP for the model $\Model$ % and the element $\AnySig_\Model$, 
with constant $\LRIPCst = (1-t)^{-1}$, metric $\LRIPMetric$, probability $1-\rho$ and error $\LRIPError=0$, as well as the BP (Def. \ref{def:BP}) with constant $\beta=1+t$, metric $\LRIPMetric$ and probability $1-\rho$.
\end{proposition}
Hence, using Theorem \ref{thm:NU-IOPLRIP}, we have shown that a reduced number of random features preserves all information when encoding signals that are (in this case) well-modeled by a union of subspaces, \emph{with respect to the associated kernel metric}. This preliminary analysis may have consequences for classical random feature bounds in a learning context \cite{Bach2015,Rudi2016}.

\section{Conclusion}
In this paper we generalized a classical property, the equivalence between the existence of an instance-optimal decoder and the LRIP, to non-linear inverse problems with possible quantization error or limited algorithmic precision, and data that live in any pseudometric set. We also formulated a version of the result for non-uniform recovery, by introducing a non-uniform version of the LRIP. To further illustrate this principle, we provided a typical proof strategy for the LRIP that one might use in practice, and gave an example of non-linear LRIP on random features for kernel approximation.

Although relatively simple in their proofs, these results may have important consequences for a large class of linear or non-linear inverse problems, where one seeks stable and robust recovery. Naturally, once the LRIP guarantees (or disproves) the existence of an instance optimal decoder, an outstanding question is the existence of \emph{efficient} algorithms that provide equivalent guarantees, as in classical compressive sensing \cite{Foucart2013} or some of its recent extensions \cite{Traonmilin2015}.

\newpage
\section*{References}
\bibliographystyle{plain}
\bibliography{library}

\newpage
\appendix

\section{Proof of Theorem \ref{thm:LRIPIOP}}\label{app:LRIPIOP}

\begin{enumerate}
\item Consider $\AnySig, \AnySig' \in \Model$. By triangular inequality we have 
\[
\IOPMetricA(\AnySig,\AnySig') \leq \IOPMetricA(\AnySig,\Decoder(\AnyOpNA,\AnyOp{\AnySig'}) + \IOPMetricA(\Decoder(\AnyOpNA,\AnyOp{\AnySig'}),\AnySig').
\]
Then, by applying the Instance Optimality Property with noise $\AnyNoise:=\AnyOp{\AnySig'}-\AnyOp{\AnySig}$ we get $\IOPMetricA(\AnySig,\Decoder(\AnyOpNA,\AnyOp{\AnySig'}) \leq \IOPCstB \normAnyMeas{\AnyOp{\AnySig'}-\AnyOp{\AnySig}} + \DecodError$, and by applying again the Instance Optimality Property it holds that $\IOPMetricA(\Decoder(\AnyOpNA,\AnyOp{\AnySig'}),\AnySig')\leq \DecodError$, hence the result.
\item Consider any signal $\tsig{\AnySig} \in \AnySigSet$ and noise $\AnyNoise\in\AnyMeasSpace$, denote $\AnyMeas = \AnyOp{\tsig{\AnySig}} + \AnyNoise$ and $\recov{\AnySig}=\Decoder(\AnyOpNA,\AnyMeas)$. Let $\AnySig_\Model \in \Model$ be any element of the model. %By application of the LRIP we have $\LRIPMetric(\AnySig_\Model,\recov{\AnySig}) \leq \LRIPCst\normAnyMeas{\AnyOp{\AnySig_\Model} - \AnyOp{\recov{\AnySig}}} + \LRIPError$ and thus:
We have:
\begin{align*}
\LRIPMetric(\tsig{\AnySig},\recov{\AnySig}) \leq&~\LRIPMetric(\tsig{\AnySig},\AnySig_\Model) + \LRIPMetric(\AnySig_\Model,\recov{\AnySig}) \\
\stackrel{\text{LRIP}}{\leq}&~\LRIPMetric(\tsig{\AnySig},\AnySig_\Model) + \LRIPCst\normAnyMeas{\AnyOp{\AnySig_\Model} - \AnyOp{\recov{\AnySig}}} + \LRIPError \\
 \leq&~\LRIPMetric(\tsig{\AnySig},\AnySig_\Model) + \LRIPCst\normAnyMeas{\AnyOp{\AnySig_\Model} - \AnyMeas} + \LRIPCst\normAnyMeas{\AnyMeas - \AnyOp{\recov{\AnySig}}} + \LRIPError~.
\end{align*}
By definition of the decoder \eqref{eq:decoder} we have $\normAnyMeas{\AnyOp{\recov{\AnySig}} - \AnyMeas} \leq \normAnyMeas{\AnyOp{\AnySig_\Model} - \AnyMeas}$ and therefore
\begin{align*}
\LRIPMetric(\tsig{\AnySig},\recov{\AnySig}) \leq&~\LRIPMetric(\tsig{\AnySig},\AnySig_\Model) + 2\LRIPCst\normAnyMeas{\AnyOp{\AnySig_\Model} - \AnyMeas} + \LRIPError \\
\leq&~\LRIPMetric(\tsig{\AnySig},\AnySig_\Model) + 2\LRIPCst\normAnyMeas{\AnyOp{\AnySig_\Model} - \AnyOp{\tsig{\AnySig}}} + 2\LRIPCst\normAnyMeas{\AnyOp{\tsig{\AnySig}} - \AnyMeas} + \LRIPError \\
\leq&~\IOPMetricB(\tsig{\AnySig},\AnySig_\Model) + 2\LRIPCst\normAnyMeas{\AnyNoise} + \LRIPError
\end{align*}
where $\IOPMetricB(\AnySig,\AnySig')=\LRIPMetric(\AnySig,\AnySig') + 2\LRIPCst\normAnyMeas{\AnyOp{\AnySig} - \AnyOp{\AnySig'}}$. Since the result is valid for all $\AnySig_\Model \in \Model$, we can take the infimum of $\IOPMetricB(\AnySig,\AnySig_\Model)$ with respect to $\AnySig_\Model \in \Model$. % and obtain the result.
\end{enumerate}

\section{Proof of Theorem \ref{thm:NU-IOPLRIP}}\label{app:NU-IOPLRIP}

Let $\tsig{\AnySig} \in \AnySigSet$ be a fixed signal and denote by $\AnySig_\Model = \Proj{\tsig{\AnySig}}\in\Model$.

%By assumption $(i)$, we 
Applying the non-uniform LRIP, with probability at least $1-\pFail_1$ on the draw of the operator $\AnyOpNA$ we have
\begin{equation}\label{eq:proofThm:NU-LRIPIOP:LRIP}
\forall \AnySig' \in \Model,~\LRIPMetric(\AnySig_\Model,\AnySig') \leq \LRIPCst \normAnyMeas{\AnyOp{\AnySig_\Model} - \AnyOp{\AnySig'_\Model}} + \LRIPError~.
\end{equation}

In the same fashion, %by assumption $(ii)$, we 
applying the non-uniform Boundedness Property, with probability at least $1-\pFail_2$ on the draw of the operator $\AnyOpNA$ we have
\begin{equation}\label{eq:proofThm:NU-LRIPIOP:BP}
\normAnyMeas{\AnyOp{\tsig{\AnySig}}-\AnyOp{\AnySig_\Model}}\leq \BPCst \BPMetric\paren{\tsig{\AnySig},\AnySig_\Model}~.
\end{equation}

Therefore, by a union bound, with probability at least $1-\pFail_1-\pFail_2$, both \eqref{eq:proofThm:NU-LRIPIOP:LRIP} and \eqref{eq:proofThm:NU-LRIPIOP:BP} are satisfied. If this is the case, for all noise vector $\AnyNoise \in \AnyMeasSpace$, denoting by $\AnyMeas = \AnyOp{\tsig{\AnySig}} + \AnyNoise$ and $\recov{\AnySig}=\Decoder(\AnyOpNA,\AnyMeas)$, it holds that:
\begin{align*}
\LRIPMetric(\tsig{\AnySig},\recov{\AnySig}) \leq&~\LRIPMetric(\tsig{\AnySig},\AnySig_\Model) + \LRIPMetric(\AnySig_\Model,\recov{\AnySig}) \\
\stackrel{\eqref{eq:proofThm:NU-LRIPIOP:LRIP}}{\leq}&~\LRIPMetric(\tsig{\AnySig},\AnySig_\Model) + \LRIPCst\normAnyMeas{\AnyOp{\AnySig_\Model} - \AnyOp{\recov{\AnySig}}} + \LRIPError \\
 \leq&~\LRIPMetric(\tsig{\AnySig},\AnySig_\Model) + \LRIPCst\normAnyMeas{\AnyOp{\AnySig_\Model} - \AnyMeas} + \LRIPCst\normAnyMeas{\AnyMeas - \AnyOp{\recov{\AnySig}}}+ \LRIPError ~.
\end{align*}
%by \eqref{eq:proofThm:NU-LRIPIOP:LRIP}. 
Once again by definition of the decoder \eqref{eq:decoder} we have $\normAnyMeas{\AnyOp{\recov{\AnySig}} - \AnyMeas} \leq \normAnyMeas{\AnyOp{\AnySig_\Model} - \AnyMeas}$ and therefore
\begin{align*}
\LRIPMetric(\tsig{\AnySig},\recov{\AnySig}) \leq&~\LRIPMetric(\tsig{\AnySig},\AnySig_\Model) + 2\LRIPCst\normAnyMeas{\AnyOp{\AnySig_\Model} - \AnyMeas}+ \LRIPError \\
\leq&~\LRIPMetric(\tsig{\AnySig},\AnySig_\Model) + 2\LRIPCst\normAnyMeas{\AnyOp{\AnySig_\Model} - \AnyOp{\tsig{\AnySig}}} + 2\LRIPCst\normAnyMeas{\AnyOp{\tsig{\AnySig}} - \AnyMeas}+ \LRIPError \\
\leq&~\IOPMetricB(\tsig{\AnySig},\AnySig_\Model) + 2\LRIPCst\normAnyMeas{\AnyNoise}+ \LRIPError.
\end{align*}
by applying \eqref{eq:proofThm:NU-LRIPIOP:BP}, which is the desired result.

\section{Proof of Proposition \ref{prop:LRIPlinear}}\label{app:LRIPlinear}

From the definition of the normalized secant set, our goal is to prove that with high probability on the draw of $\AnyOpNA$, for all $\mathbf{s}\in\secant$ we have $\normAnyMeas{\AnyOpNA\mathbf{s}}\geq 1-t$.

Let $\coveps,\varepsilon>0$ be small constants which values we shall define later and $N := \covnum{\secant}{\IOPMetricA}{\coveps}$ be the covering numbers of the normalized secant set. Let $\mathbf{s}_1,\ldots,\mathbf{s}_N$ be an $\coveps$-covering of the normalized secant set $\secant$. %, denoting by $\AnySig_i,\AnySig'_i\in\Model$ the elements of the model such that $\mathbf{s}_i = \frac{\AnySig_i - \AnySig'_i}{\normAnySig{\AnySig_\Model,\AnySig'_i)}}$ and $0<\LRIPMetric(\AnySig_\Model,\AnySig'_i)\leq \varepsilon$.

By the concentration property, it holds that, with probability at least $1-Ne^{-c(t/2)}$, for all $\mathbf{s}_i$ we have 
\begin{equation}\label{eq:linLRIPproofpointwise}
\normAnyMeas{\AnyOpNA\mathbf{s}_i} \geq 1-t/2
\end{equation}
%Our goal is to extend this property to any element $\mathbf{s}$ in the model $\Model$. Let $\AnySig\in\Model$ be any element of the model. We distinguish two cases.

Now, given any element of the normalized secant set $\mathbf{s}\in\secant$, one can find an element of the covering $\mathbf{s}_i$ such that $\normAnySig{\mathbf{s} - \mathbf{s}_i}\leq \coveps$. Assuming \eqref{eq:linLRIPproofpointwise} holds, we have
\begin{align*}
\normAnyMeas{\AnyOpNA\mathbf{s}} \geq \normAnyMeas{\AnyOpNA\mathbf{s_i}} - \normAnyMeas{\AnyOp{\mathbf{s}-\mathbf{s_i}}} \geq 1- \frac{t}{2} - C\coveps
\end{align*}
and therefore by choosing $\coveps = t/(2C)$  we obtain the desired result.

\section{Proof of Proposition \ref{prop:nonlinLRIP}}\label{app:nonlinLRIP}

Fix any $\AnySig_\Model \in\Model$.
Let $\coveps,\coveps',\varepsilon>0$ be small constants which values we shall define later, such that $\varepsilon\leq \varepsilon_0$. Define $\Model_\varepsilon = \Model\backslash \set{\AnySig \in \Model,~\LRIPMetric(\AnySig_\Model,\AnySig)\leq\varepsilon}$ the model with a ball around $\AnySig_\Model$ removed,
and $N := \covnum{\Model_\varepsilon}{\LRIPMetric}{\coveps}$, $N' := \covnum{\secant_\varepsilon}{\LRIPMetric}{\coveps'}$. Let $\AnySig_1,\ldots,\AnySig_N$ and $\mathbf{s}_1,\ldots,\mathbf{s}_{N'}$ be, respectively, a $\coveps$-covering of $\Model_\varepsilon$ and a $\coveps'$-covering of $\secant_\varepsilon$, with $\AnySig'_i\in\Model$ defined such that $\mathbf{s}_i = \frac{\AnySig_\Model - \AnySig'_i}{\LRIPMetric(\AnySig_\Model,\AnySig'_i)}$ and $0<\LRIPMetric(\AnySig_\Model,\AnySig'_i)\leq \varepsilon$.

By the concentration property, it holds that, with probability at least $1-(N+N')e^{-c(t/2)}$, for all $\tilde\AnySig = \AnySig_i$ or $\tilde\AnySig = \AnySig'_i$ for all indices $i$, we have 
\begin{equation}\label{eq:nonlinLRIPproofpointwise}
\normAnyMeas{\AnyOp{\AnySig_\Model} - \AnyOp{\tilde\AnySig}} \geq (1-t/2) \cdot \LRIPMetric(\AnySig_\Model,\tilde\AnySig).
\end{equation}
Our goal is to extend this property to any element $\tilde{\AnySig} = \AnySig$ in the model $\Model$.

Let $\AnySig\in\Model$ be any element of the model. We distinguish two cases.
If $\AnySig\in\Model_\varepsilon$, $i.e.$ $\LRIPMetric(\AnySig_\Model,\AnySig)\geq \varepsilon$, we consider $\AnySig_i \in \Model_\varepsilon$ an element of the covering of $\Model_\varepsilon$ such that $\LRIPMetric(\AnySig, \AnySig_i)\leq \coveps$. We have:
\begin{align*}
\normAnyMeas{\tfrac{\AnyOp{\AnySig_\Model} - \AnyOp{\AnySig}}{\LRIPMetric(\AnySig_\Model,\AnySig)} - \tfrac{\AnyOp{\AnySig_\Model} - \AnyOp{\AnySig_i}}{\LRIPMetric(\AnySig_\Model,\AnySig_i)}} \leq&~
\normAnyMeas{\tfrac{\AnyOp{\AnySig_\Model} - \AnyOp{\AnySig}}{\LRIPMetric(\AnySig_\Model,\AnySig)} - \tfrac{\AnyOp{\AnySig_\Model} - \AnyOp{\AnySig_i}}{\LRIPMetric(\AnySig_\Model,\AnySig)}} + \normAnyMeas{\tfrac{\AnyOp{\AnySig_\Model} - \AnyOp{\AnySig_i}}{\LRIPMetric(\AnySig_\Model,\AnySig)} - \tfrac{\AnyOp{\AnySig_\Model} - \AnyOp{\AnySig_i}}{\LRIPMetric(\AnySig_\Model,\AnySig_i)}} \\
=&~\tfrac{\normAnyMeas{\AnyOp{\AnySig} - \AnyOp{\AnySig_i}}}{\LRIPMetric(\AnySig_\Model,\AnySig)} + \normAnyMeas{\AnyOp{\AnySig_\Model} - \AnyOp{\AnySig_i}}\abs{\tfrac{1}{\LRIPMetric(\AnySig_\Model,\AnySig)} - \tfrac{1}{\LRIPMetric(\AnySig_\Model,\AnySig_i)}} \\
\stackrel{(i),(ii)}{\leq}&~\tfrac{C_1 \coveps}{\varepsilon} + C_1 M_\Model \tfrac{\LRIPMetric(\AnySig,\AnySig_i)}{\LRIPMetric(\AnySig_\Model,\AnySig)\LRIPMetric(\AnySig_\Model,\AnySig_i)} \\
\leq&~\tfrac{C_1 \coveps}{\varepsilon}\paren{1+\tfrac{M_\Model }{\varepsilon}}
\end{align*}
and therefore
\begin{equation}
\label{eq:nonlinLRIPproofbound1}
\normAnyMeas{\frac{\AnyOp{\AnySig_\Model} - \AnyOp{\AnySig}}{\LRIPMetric(\AnySig_\Model,\AnySig)}} \geq 1-t/2-\frac{C_1 \coveps}{\varepsilon}\paren{1+\frac{M_\Model}{\varepsilon}}
\end{equation}

Now, when $\LRIPMetric(\AnySig_\Model,\AnySig)\leq \varepsilon$, we define $\mathbf{z} := \frac{\AnySig_\Model-\AnySig}{\LRIPMetric(\AnySig_\Model,\AnySig)}$ and note that $\mathbf{z}\in\secant_\varepsilon$. We approximate it by an element of the covering of the normalized secant set $\mathbf{s}_i = \frac{\AnySig_\Model - \AnySig'_i}{\LRIPMetric(\AnySig_\Model,\AnySig'_i)}\in\secant_\varepsilon$ (meaning that $\LRIPMetric(\AnySig_\Model,\AnySig'_i) \leq \varepsilon$) that verifies $\LRIPMetric(\mathbf{z}, \mathbf{s}_i) \leq \coveps'$. Then, we have
\begin{align*}
\normAnyMeas{\tfrac{\AnyOp{\AnySig_\Model} - \AnyOp{\AnySig}}{\LRIPMetric(\AnySig_\Model,\AnySig)} - \tfrac{\AnyOp{\AnySig_\Model} - \AnyOp{\AnySig'_i}}{\LRIPMetric(\AnySig_\Model,\AnySig'_i)}} \leq&~
\normAnyMeas{\tfrac{\AnyOp{\AnySig_\Model} - \AnyOp{\AnySig}}{\LRIPMetric(\AnySig_\Model,\AnySig)} - D_\AnyOpNA\mathbf{z}} + \normAnyMeas{D_\AnyOpNA(\mathbf{z} - \mathbf{s}_i)} + \normAnyMeas{D_\AnyOpNA\mathbf{s}_i - \tfrac{\AnyOp{\AnySig_\Model} - \AnyOp{\AnySig_i}}{\LRIPMetric(\AnySig_\Model,\AnySig_i)}} \\
\stackrel{(iv)}{\leq}&~C_2 \LRIPMetric(\AnySig_\Model,\AnySig) + C_3 \coveps'+ C_2\LRIPMetric(\AnySig_\Model,\AnySig'_i) \leq 2C_2 \varepsilon + C_3 \coveps'
\end{align*}
and therefore
\begin{equation}
\label{eq:nonlinLRIPproofbound2}
\normAnyMeas{\frac{\AnyOp{\AnySig_\Model} - \AnyOp{\AnySig}}{\LRIPMetric(\AnySig_\Model,\AnySig)}} \geq 1-t/2-(2C_2 \varepsilon + C_3 \coveps')
\end{equation}

To conclude, we set $\varepsilon \leq \min\paren{\varepsilon_0,\frac{t}{8C_2}}$, $\coveps' \leq \frac{t}{4C_3}$, $\coveps \leq \frac{t\varepsilon^2/(4C_1)}{\varepsilon + M_\Model}$ to obtain the desired result.

\section{Proof of Proposition \ref{prop:rf}}\label{app:rf}

\paragraph{Concentration property.} The concentration result is based on the fact that $\norm{\AnyOp{\AnySig} - \AnyOp{\AnySig}}_2^2 \approx \Exp_\freq \abs{\varphi_\freq(\AnySig) - \varphi_\freq(\AnySig')}^2 = \LRIPMetric(\AnySig,\AnySig')^2$ by definition of the random features. Using simple function studies and a Bernstein concentration inequality with a control on moments of all orders, it is possible to show (see \cite{Gribonval2017}, eq. (160) then Prop. 6.11) that the concentration result \eqref{eq:concentration_linear} is valid with $c(t) \propto \frac{m t^2}{1+t}$ (we do not reproduce the detailed proof here for brevity). 

Since this Berstein inequality is in fact valid for all vectors (not necessarily in the model), as a consequence the Boundedness Property is also satisfied with constant $\beta=1+t$, metric $\LRIPMetric$ and probability $1-e^{-c(t)}$.

We now check hypotheses $(i)$--$(iv)$ in Proposition \ref{prop:nonlinLRIP}. We are going to repeatedly use the fact that for $\AnySig,\AnySig'\in\Model$, since $\norm{\AnySig}_2\leq M$, we have
\begin{equation}\label{eq:rfnorm}
\ell \norm{\AnySig - \AnySig'}_2 \leq \LRIPMetric(\AnySig,\AnySig') \leq L \norm{\AnySig - \AnySig'}_2
\end{equation}
where $\ell := 2\cdot \frac{1-\exp\paren{-M^2/(2\sigma^2)}}{M}$ and $L = \frac{M}{\sigma^2}$.
\paragraph{(i)} Using a first-order Taylor expansion we have $\norm{\AnyOp{\AnySig} - \AnyOp{\AnySig'}}_2 \leq \paren{\sum_{j=1}^m \frac{\norm{\freq_j}_2^2}{m f(\freq_j)^2}}^\frac12\norm{\AnySig - \AnySig'}_2\lesssim \sqrt{\gamma_2} \norm{\AnySig - \AnySig'}_2$ hence by \eqref{eq:rfnorm} hypothesis $(i)$ is valid with $C_1 \propto \frac{\sqrt{\gamma_2}}{\ell}$.

\paragraph{(ii)} It is immediate that $M_\Model = ML$. Then, using the well-known fact that for each $S_j$ we have $\covnum{S_j\cap B_M}{\norm{\cdot}_2}{\coveps} = \order{\paren{\frac{M}{\coveps}}^s}$, by a union bound and \eqref{eq:rfnorm} we have $\covnum{\Model}{\LRIPMetric}{\coveps} \leq \order{N\paren{\frac{LM}{\coveps}}^s}$.

\paragraph{(iii)} Similar to the model, for all $\varepsilon$ the normalized secant set is included in a union of subspace $\secant_\varepsilon \subset \bigcup_{i,j=1}^N S_{ij} \cap B_{M/\ell}$, with norm controlled by $M/\ell$ by \eqref{eq:rfnorm}, and where $S_{ij} = S_i + S_j$ (sum of subspaces). Hence, since $\covnum{S_{ij}\cap B_{M/\ell}}{\norm{\cdot}_2}{\coveps} = \order{\paren{\frac{M}{\ell\coveps}}^{2s}}$, by a union bound and \eqref{eq:rfnorm} we have $\covnum{\secant_\varepsilon}{\LRIPMetric}{\coveps} \leq \order{N^2\paren{\frac{LM}{\ell\coveps}}^{2s}}$.

\paragraph{(iv)} Finally, by a Taylor expansion we have
\[
\varphi_\freq(\AnySig) - \varphi_\freq(\AnySig') = \frac{\imaginaryi\freq^\top}{f(\freq)}(\AnySig - \AnySig') e^{\imaginaryi\freq^\top \AnySig} - \frac{1}{2}\cdot\frac{(\freq^\top(\AnySig-\AnySig'))^2}{f(\freq)}e^{\imaginaryi\freq^\top \AnySig''}
\]
where $\AnySig'' = \theta \AnySig + (1-\theta)\AnySig'$. Hence hypothesis $(iv)$ is satisfied with $C_2 = C_1$ and $C_3 = \frac{\sqrt{\gamma_4}}{\ell^2}$, which concludes the proof.
\end{document}